\documentclass[twocolumn,showpacs,preprintnumbers,amsmath,amssymb]{revtex4}

\usepackage{amsmath}    % need for subequations
\usepackage{graphicx}   % for figures

\newcommand{\be}{\begin{eqnarray}}
\newcommand{\ee}{\end{eqnarray}}
\newcommand{\bdm}{\begin{displaymath}}
\newcommand{\edm}{\end{displaymath}}
\newcommand{\nn}{\nonumber}

\begin{document}

\title{Relativistic reactions visualized through right triangles in space}
\author{Theocharis A.~Apostolatos}
\affiliation{Section of Astronomy, Astrophysics, and Mechanics\\
University of Athens\\ Panepistimiopolis, Zografos, GR-15783,
Athens, Greece }
 \email{thapostol@phys.uoa.gr}   %optional
\date{\today}

\begin{abstract}
In this article we exploit the fact that the special relativistic formula which
relates the energy and the 3-momentum of an elementary particle with its
rest mass, resembles the pythagorean theorem for right triangles. Using such triangles,
suitably arranged, we can prove all kind of equalities or inequalities
concerning the kinematical properties of elementary particles
in a wide variety of cases regarding particles' collision, decay or production.
Moreover, relations that are somehow hard to
produce by the usual analytic methods arise much more naturally
through geometric constructions based on right triangles.
\end{abstract}

\maketitle

%%%%%%%%%%%%%%%%%%%%%%%
\section{Introduction}
\label{sec:1}
%%%%%%%%%%%%%%%%%%%%%%%

After the notorious Address delivered by H. Minkowski at the Assembly
of German Natural Scientists and Physicians
in 1908 \cite{Mink1908},  which was devoted to space and time,
all physical quantities which were previously described as
vectors in 3-D Euclidean space, were recast accordingly to form either vectors
in a 4-D pseudo-Euclidean space-time (like the 4-momentum),
or to tensorial objects   (like the electromagnetic field).
%Especially momentum, a physical quantity of great importance in
%collisions, turned into 4-momentum, with energy playing the role of its time-component and 3-momentum corresponding to its spatial
%part.
The magnitude of all 4-vectors that are used in special relativity is
defined as
\[
||A||^2=A^\mu A_\mu=-(A^0)^2+\textbf{A}^2,
\]
where $\textbf{A}$ is the 3-D vector that represents the spatial part of $A^\mu$.
This expression resembles the pythagorean theorem relating the sides of a
right triangle, with the time component $A^0$ playing the role of
triangle's hypotenuse. This property has been extensively used in
many introductory textbooks devoted to Special
Relativity \cite{TaylorWheeler}, in order to visualize
relations that apply between various physical quantities which become interrelated in
the framework of Minkowski's space-time. However, there is an intrinsic problem in
drawing such right triangles in a generic situation: one has to draw
4-D objects since one of the sides of the triangle is already a vector living in the 3-D Euclidean space,
while the other perpendicular side should extend beyond this 3-D space. This practical problem arises
whenever one deals with more than two such triangles, since then there is not
enough space to draw everything. In all cases explored here we will show that
the most generic situation can always be analyzed with only two such triangles,
the vector sides of which are defining a plane while the third dimension could be used to
develop the other perpendicular sides of the corresponding triangles.

In this paper we demonstrate the use of such right
triangles in analyzing elementary-particle reactions;  the
corresponding sides of each triangle will represent the energy, $E$, the
3-momentum,  $\textbf{p}$, and the rest mass, $m$, of each particle
(we will work with units where the velocity of light
is $c=1$). This diagrammatic tool 
has hardly been used in textbooks (for a recent article that refers
to this tool see \cite{Okun})
for these physical quantities, although, as it will be shown
later, the quantitative use of such geometric tools is in many cases
 advantageous compared to the usual analytic
 methods. Solving problems  concerning particles that collide with
 each other and produce a number of new particles by the usual algebraic
 way is quite often a rather complicated procedure, mainly due to the
 fact that apart of energy and momentum conservation there are also
 constraints of quadratic form from the energy-momentum-mass relation. Thus if
 one does not follow the most clever way to analyze the problem,
 computations may become really meshy. The situation looks like a labyrinth;
 if one does not know the right path connecting the starting point with the endpoint,
 various paths leading to dead ends will be chosen until the aim is accomplished.

 By using right triangles for each
 particle, arranged in space in a convenient way,  we can get answers about
 all physical quantities in a much more straightforward fashion.
 Moreover, these geometric constructions can easily be used by someone
 to draw preliminary qualitative conclusions, like ``is a reaction allowed or is it
 forbidden'', or ``could the angle between the lines of motion of two products exceed
 $90^\circ$ or not?'' Also, if one intends to construct a new problem,
 the graphical analysis through right triangles may help in posing well defined
 questions that have a clear answer.
 The only difficulty in this geometrical
 approach of solving or sketching
 problems is to draw a suitable 3-D construction of right
 triangles. We will demonstrate the power of this geometrical
 method by solving accordingly  a few characteristic problems.
 For each problem we will draw the right triangles in such an
 arrangement in 3-D space that answers will come out quickly and easily.

%%%%%%%%%%%%%%%%%%%%%%%%%%%%%%%%%%%%%%%%%%%%%%
 \section{Orthogonal triangles corresponding to particles}
 \label{sec:2}
%%%%%%%%%%%%%%%%%%%%%%%%%%%%%%%%%%%%%%%%%%%%%%

\begin{figure}[h]
\begin{center}
\centerline{\includegraphics[width=13pc] {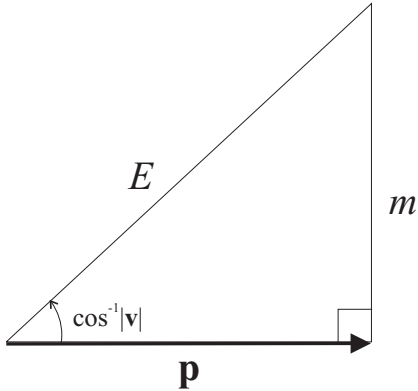}}
%\scalebox{0.50}{\includegraphics{fig1.eps}}
\caption{\label{fig:1} This is the basic geometric unit that will be used throughout
the paper to obtain geometric solutions in a wide variety of problems
related to relativistic collisions. It is a right triangle that represents the
relativistic relation $E^2=\textbf{p}^2+m^2$. For the two extreme cases (classical particle
and highly relativistic particle) the corresponding triangles are like degenerate
two-sided triangles. For a photon-like particle  one
of the sides of the right triangle has exactly zero length.}
\end{center}
\end{figure}

As was mentioned previously the basic unit in our geometric construction
will be a right triangle with hypotenuse $E$, and two perpendicular sides
with magnitude $|\textbf{p}|$ and $m$ respectively, where $E,\textbf{p},m$ are the energy,
the 3-momentum, and the mass of a single particle (see Fig.~\ref{fig:1}), all of them measured in
a specific inertial frame of reference. Since $\textbf{p}$ is a vector in
a 3-D Euclidean space, the corresponding side will be represented by a vector;
hence particular attention should often be paid in arranging the vectorial side of the triangle along the
corresponding direction in a 3-D drawing. The kinetic energy of a particle is defined as
\be
T=E-m,
\ee
thus it is represented by the difference between the corresponding
sides of the right triangle. Also the velocity of the particle in the
particular frame of reference is given by the ratio
\be
\mathbf{v}=\frac{\mathbf{p}}{E},
\ee
that is by the cosine of the right angle  subtending the mass side of the triangle.

%%%%%%%%%%%%%%%%%%%
%%%%%%%%%%%%%%%%%%%

The diagram corresponding to the two extreme cases, that of a classical particle and that
of a highly relativistic one, should be considered separately.
The former one will be represented by a degenerate right triangle with its
hypotenuse,  $E$, being almost equal to the $m$ side, while the latter particle will be represented again
by a degenerate right triangle
with its hypotenuse, $E$, being almost equal to the $|\textbf{p}|$ side. By Taylor expansion we see
that in the former (classical) case
\be
E=\sqrt{m^2+\textbf{p}^2}=m \sqrt{1+\frac{\textbf{p}^2}{m^2}}\cong m+ \frac{\textbf{p}^2}{2 m}, %\cong m+ \frac{\textbf{p}^2}{2 E}
\ee
the last term being the classical Newtonian kinetic energy,
while in the latter (ultra-relativistic) case the corresponding relation is
\be
E \cong |\textbf{p}|+ \frac{m^2}{2 |\textbf{p}|}%\cong |\textbf{p}|+ \frac{m^2}{2 E}
\ee
by mere similarity of the corresponding sides ($m \leftrightarrow |\textbf{p}|$) in these two extreme cases. Of course photons, or any other
zero rest-mass particle will be represented by a degenerate right triangle with two equal sides ($E=|\textbf{p}|$).

At this point we should note that the shape of any triangle depends on the frame of reference on which the
specific particle is observed (two of its sides $E,|\textbf{p}|$ are frame dependent).
Therefore a geometric construction of many such right triangles at the same diagram will refer
to a single frame of reference for all particles. Although in algebraic analysis of problems related to
relativistic collisions we usually change frame of reference to make computations easier, there is no
need to appeal to such tricks in geometric solutions, since the geometric conclusion will be clear, independently of
the frame of reference used to draw all triangles. Actually by avoiding Lorentz transformations to shift
frames of reference we avoid a lot of algebraic computations.

The geometric solutions are based on conservation of 4-momentum in
relativistic reactions; that is simultaneous conservation of the total energy, and the total 3-momentum of all particles
that participate in a reaction. Therefore all geometric constructions that correspond to allowed
arrangements for the kinematical characteristics of the particles involved, share common total length of $E$-sides
and vectorial sum of $\textbf{p}$ sides.

Next, we proceed to describe some properties regarding the extremum of specific quantities
for a number of particles. These general propositions, that we will prove graphically,
will be later used in some of the problems that we will analyze by means of our geometrical method.

\noindent\textbf{Proposition I:} The total 3-momentum of $N$ particles with energies $E_i$ and masses $m_i$,
respectively ($i=1,2,\ldots,N$), gets maximized if all particles are moving along the same direction.

\noindent\textbf{Proof:} This proposition is quite obvious, and there is no need to use any right triangles to prove it.
However the proof is purely geometrical, as in the forthcoming propositions and problems.
Given the magnitudes of all energies and masses, the
magnitudes of all 3-momenta are fixed:
$|\textbf{p}_i|=\sqrt{E_i^2-m_i^2}$. Therefore if we connect all
these vectors head-to-tail we construct the total 3-momentum of the
system of particles (which could be either the reacting particles or
the products). From the triangle inequality the
total 3-momentum of all particles $\textbf{p}_\textrm{tot}$ satisfies the
inequality
\be
|\textbf{p}_{\textrm{tot}}| \leq
|\textbf{p}_1|+|\textbf{p}_2|+\ldots+|\textbf{p}_N|,
\ee
where the
equality holds if $\textbf{p}_i/|\textbf{p}_i| =  \textbf{p}_j/|\textbf{p}_j|$  for all $i,j=1,2,\ldots,N$,
that is when all the 3-momenta are aligned to each other (common direction of motion for all particles).
Intuitively we could just say that the head of the last 3-momentum vector and the tail of the first one
are further apart when the broken line of the corresponding vectors is straightened.
%%%%%%%%%%%%%%%%%%%%%%%%%

\noindent\textbf{Proposition II:} Two or more particles can be considered as a single particle with respect to
geometric representation by right triangles.

\noindent\textbf{Proof:} It is obvious that we could always construct a right triangle with one leg equal to the total
3-momentum of the particles
\be
{\bf  p}_{\textrm{tot}}=\sum_i {\bf  p}_i
\ee
and hypotenuse equal to the sum of the energies of all particles
\be
E_{\textrm{tot}}=\sum_i E_i
\label{totAne}
\ee
since for each particle it holds $E_i \geq |{\bf  p}_i|$ (the equality holds only for massless
particles). The latter sum (Eq.~(\ref{totEne})) could serve as a hypotenuse since
\be
E_{\textrm{tot}}=\sum_i E_i \geq \sum_i |{\bf  p}_i| \geq \left| \sum_i {\bf  p}_i \right| =|{\bf  p}_{\textrm{tot}}|.
\ee
It should be noted though that this representation is not one-to-one; for a specific
value of $E_{\textrm{tot}}$ and ${\bf  p}_{\textrm{tot}}$
there is too much freedom in choosing the energy and 3-momentum of each particle.

The characteristic mass
of the representative right triangle is not equal to the sum of the corresponding particles' masses;
it is equal to
$(E_\textrm{tot}^2-{\bf  p}_\textrm{tot}^2 )^{1/2}$,
which in its turn is equal to the total  energy in the center-of-momentum frame for these particles
(the frame in which ${\bf  p}_{\textrm{tot}}=0$.
In propositions III and IV we will actually replace many particles in the corresponding inductive proofs
by a single one. As we shall see in both cases the minimization/maximization procedure leads to equal velocities
for all particles. This is a distinct
representation of many particles by a single composite particle. It is the only case where the representation is one-to-one
since then the rest mass in the center-of-momentum frame is exactly equal to the some of all rest masses.
From the point of view of our geometrical method there is only one way to construct many similar right triangles
with one of their sides (the mass sides) given; this corresponds to particles with equal velocities.
%%%%%%%%%%%%%%%%%%%%%%%%%%%%%%

\noindent\textbf{Proposition III:} For a given total 3-momentum of $N$ particles, the sum of
their  energies $\sum_i E_i$ assumes its minimum value when all particles move with the same velocity $\textbf{v}$.

\begin{figure}[h]
\begin{center}
\centerline{\includegraphics[width=19pc] {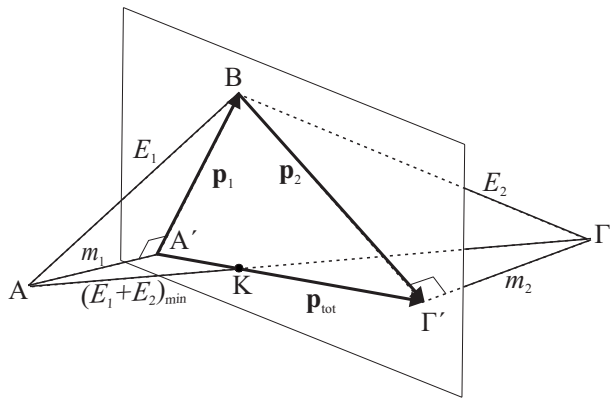}}
%\scalebox{0.60}{\includegraphics{fig2.eps}}
\caption{\label{fig:2} In order to minimize the total energy of two particles
we should split their total 3-momentum in two parts with ratio equal to the ratio of their masses (point B
should be moved to point K).
The mass segments have been drawn perpendicular to the plane of momenta screen.}
\end{center}
\end{figure}

\noindent\textbf{Proof:} Let us begin assuming that there are only two particles. Consider the geometric
construction of Fig.~\ref{fig:2} depicting two right triangles with sides
$E_1,m_1,\textbf{p}_1$, and $E_2,m_2,\textbf{p}_2$. From now on we will call the plane defined by the
vector sides $\textbf{p}_1,\textbf{p}_2$ \textit{momenta screen}, since we are going
to use extensively this plane to move the vectors of 3-momenta around.
On this plane we have also drawn the total 3-momentum of the two particles,
which according to Proposition III is assumed to be fixed. Any point B on momenta screen
represents a specific split for the 3-momenta of the two particles.
When two line segments, $\textrm{AA}^\prime$ and $\Gamma\Gamma^\prime$ with lengths equal
to the masses of the two particles are drawn perpendicular to the
plane of momenta screen as in Fig.~\ref{fig:2}, the total length of the
hypotenuses $\textrm{AB}$, $\textrm{B}\Gamma$ corresponds to the total energy
of the particles. Clearly the shortest distance between the two points
$\textrm{A}$ and $\Gamma$ (which corresponds to the minimum total energy)
will be the straight distance $\textrm{A}\Gamma$. This straight line
intersects vector $\textbf{p}_\textrm{tot}$ at point $\textrm{K}$, forming two similar right triangles
$\textrm{AA}^\prime\textrm{K}$ and $\Gamma\Gamma^\prime\textrm{K}$. From
similarity of the two triangles the equality of the ratios of masses and 3-momenta is
apparent.

By induction we could generalize our result to
more than two particles:
For $N$ particles we represent the first $N-1$ particles by a single right triangle according to
Proposition II and we minimize the energy of the $N$-th particle and the rest particles.
Then we proceed to minimize the energy of the first $N-1$ particles and so on.
Thus we conclude that the
total energy of all particles is minimized when the total 3-momentum is split in $N$ segments
proportional to the corresponding masses of the particles. Then the right triangles that correspond to all particles
are similar triangles and thus all particles should have the same velocity (cf.~Fig.~\ref{fig:1}).

Note that in the case where three or more particles are
involved we cannot draw any momenta screen and then construct all the corresponding right
triangles perpendicular to such a plane, since three or
more arbitrary vectors do not lie in the same plane. This explains why we had to appeal to
induction to prove this proposition.

%%%%%%%%%%%%%%%%%%%
%%%%%%%%%%%%%%%%%%%

At this point we could add a bit of information to Proposition II.
According to Proposition III the mass of the composite particle described
by a single right triangle, which represents a number of particles (see Proposition II),
\be
M_{\textrm{comp}}= \sqrt{ E_{\textrm{tot}}^2-{\bf p}_{\textrm{tot}}^2 },
\label{Mcomposite}
\ee
is higher than the total rest mass of the specific particles, since for a given
${\bf p}_{\textrm{tot}}|$ of such particles their total energy is minimized
when all particles are moving with the same velocity.
As it was noticed in the analysis of Proposition II, this optimum case corresponds to
$M_{\textrm{comp}} = \sum_i m_i$.
Thus
\be
M_{\textrm{comp}} \geq \sqrt{ (E_{\textrm{tot}}\left.\right|_{\min})^2-{\bf p}_{\textrm{tot}}^2 } =
\sum_i m_i.
\ee

%%%%%%%%%%%%%%%%%%%%%%

\noindent\textbf{Proposition IV:} For a given total energy of $N$ particles the sum of
their  3-momenta $\sum_i \textbf{p}_i$ assumes its maximum magnitude if all
particles are moving with the same velocity $\mathbf{v}$.
% 3-momenta are
%parallel to each other, having all the same direction, and the ratio of magnitudes between any two of those
%is equal to the corresponding ratio of their masses ($\textbf{p}_i/m_i=\textbf{p}_j/m_j$).
This is simply the inverse of Proposition III.

\begin{figure}[h]
\begin{center}
\scalebox{0.50}{\includegraphics{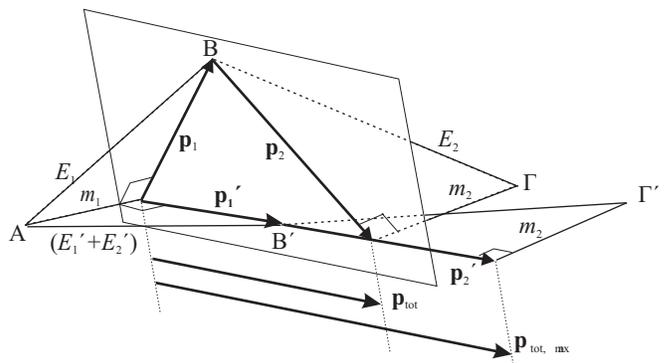}}
%\centerline{\includegraphics[width=20pc] {fig3.eps}}
\caption{\label{fig:3} The procedure of maximizing the magnitude of the total 3-momentum of two particles for a given
total energy. Initially the two right triangles are not coplanar. By moving the mass segments $m_1,m_2$
around we maximize the total 3-momentum by putting them as far apart as possible. Then two similar right triangles
form. This situation corresponds to the same velocity for both particles.}
\end{center}
\end{figure}

\noindent\textbf{Proof:} We shall use induction once again. Referring to Fig.~\ref{fig:3} that
depicts the right triangles of two particles, it is clear that we are free to move the mass segments $m_1,m_2$
(which stick out of the plane of momenta screen in a perpendicular fashion) around as long as we keep the sum of
the two hypotenuses,  $E_1+E_2$, represented by the length of segments $\textrm{AB}$ and $\textrm{B}\Gamma$,
fixed. The total 3-momentum of the two particles is the distance between the projections of points $\textrm{A}$ and $\textrm{B}$
on momenta screen. In order to render this distance maximum we have to move the two mass segments as far as possible
from each other. The maximum distance between the mass segments, compatible with the
total energy constraint, is
achieved when the line $\textrm{AB}\Gamma^\prime$ is a straight line. Then the two right triangles that  form
are similar, and the ratio of the two masses $m_1/m_2$ equals the ratio of the magnitudes of the two 3-momenta
$|\mathbf{p}'_1|/|\mathbf{p}'_2|$, while both 3-momenta lie along the same line. This configuration corresponds again
to common velocity for both particles. For more than two particles, we first maximize the 3-momentum of the first pair of particles
while keeping the sum of their energies constant. Then we replace these two particles by a single one
which moves with the common velocity of the two particles and has mass equal to the sum of the masses of the initial particles
and then we proceed further. In every step we maximize the total 3-momentum, without changing the total energy,
by attributing the same velocity to all particles considered up to that point.

%%%%%%%%%%%%%%%%%%%
%%%%%%%%%%%%%%%%%%%
%\end{document}

As mentioned before, the similarity of triangles that arises
from the graphical solution of the last two propositions corresponds to the case where all
particles are moving with the same velocity. This is also the velocity of the center-of-momentum frame.
In this frame all particles are then at rest. Therefore this minimization/maximization is succeeded
when the particles are still in their center of momentum.

The following problems of relativistic reactions will be presented in a sequence of gradual
difficulty with respect to the corresponding analytic solutions for each of them. On the other hand applying
the geometrical tools in order to  draw, at least qualitative conclusions, is in all cases of the same difficulty
and much more direct.
Thus, although in the first couple of problems one may argue that the geometric method does not
offer an easier solution than the
usual algebraic solution, for the rest of problems the directness of the geometrical proof is apparent.

%%%%%%%%%%%%%%%%%%%%%%%%%%%%%%%%%%%%%%%%%%%%%%
\section{Problem 1}
\label{sec:3}
%%%%%%%%%%%%%%%%%%%%%%%%%%%%%%%%%%%%%%%%%%%%%%

We start familiarizing ourselves with the geometric constructions by treating accordingly the following
simple problem: Show that a photon cannot be split spontaneously  into two or more massive particles (like in pair production).

%%%%%%%%%%%%%%%%%%%
\begin{figure}[h]
\begin{center}
%\scalebox{0.50}{\includegraphics{fig4.eps}}
\centerline{\includegraphics[width=20pc] {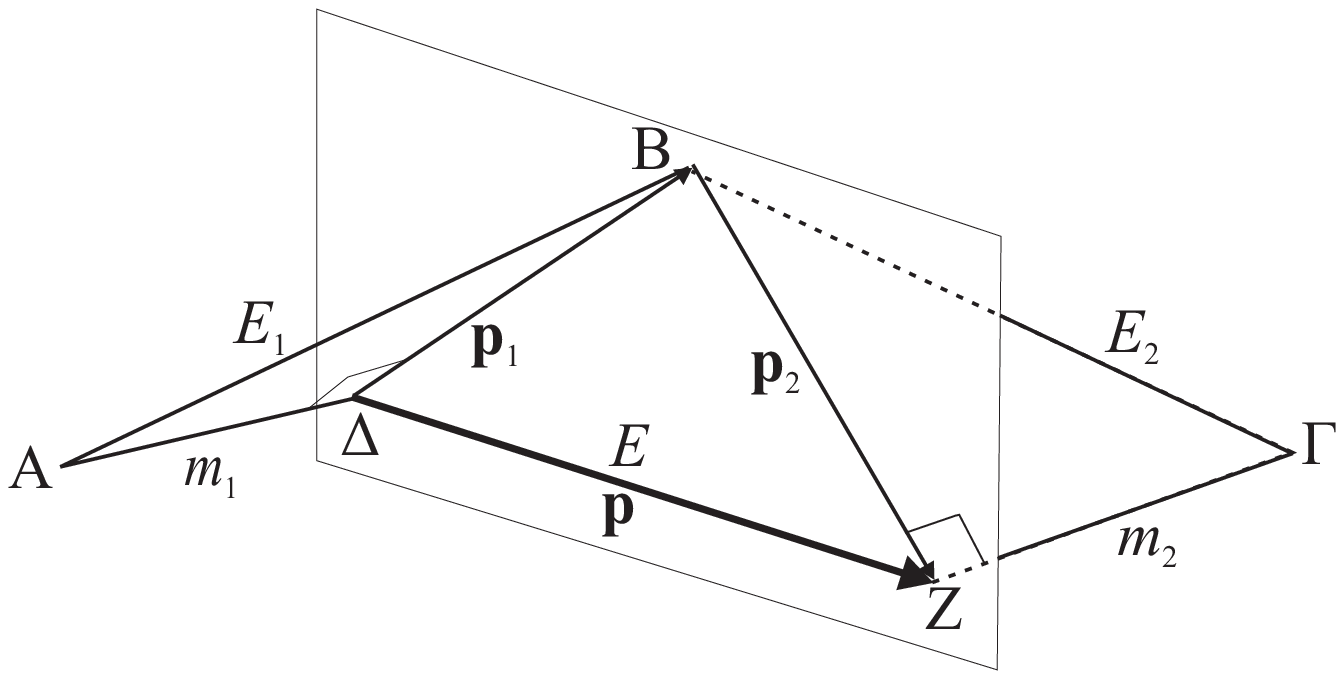}}
\caption{\label{fig:4} The geometric construction for the
spontaneous decay of a photon.}
\end{center}
\end{figure}
%%%%%%%%%%%%%%%%%%%

In Figure \ref{fig:4} we have drawn on momenta screen the initial 3-momentum ${\bf p}$
of the photon the magnitude of which equals the energy of the photon $E$. The
3-momenta of two new hypothetic particles have also been drawn on momenta screen, as well as their energies
coming out of this plane. However, the following
inequality
\be
E_1+E_2=\textrm{AB}+\textrm{B}\Gamma \geq \textrm{A}\Gamma > \Delta\textrm{Z}=E
\ee
is true. Therefore, due to conservation of energy this type of reaction is forbidden.
The same is also true for more than two particles, since all but one particles could have been replaced by
a single representative triangle according to Proposition II.

The case of a single massive particle produced by a photon is obviously forbidden
since a degenerate triangle (photon) could not be equal to a non-degenerate triangle (massive particle).

%%%%%%%%%%%%%%%%%%%%%%%%%%%%%%%%%%%%%%%%%%%%%%
\section{Problem 2}
\label{sec:4}
%%%%%%%%%%%%%%%%%%%%%%%%%%%%%%%%%%%%%%%%%%%%%%

Next, we show that  particle creation from a single photon is allowed,
as long as a massive particle (usually an atomic nucleus), assumed to be initially motionless,
is used to interact with the photon, while after the reaction this particle remains unaltered.
Therefore the massive particle plays the role of
a catalyst for such a reaction. %while this particle remains the same after the reaction.
We will also obtain the sufficient condition for this reaction to happen.

%%%%%%%%%%%%%%%%%%%
\begin{figure}[h]
\begin{center}
\scalebox{0.60}{\includegraphics{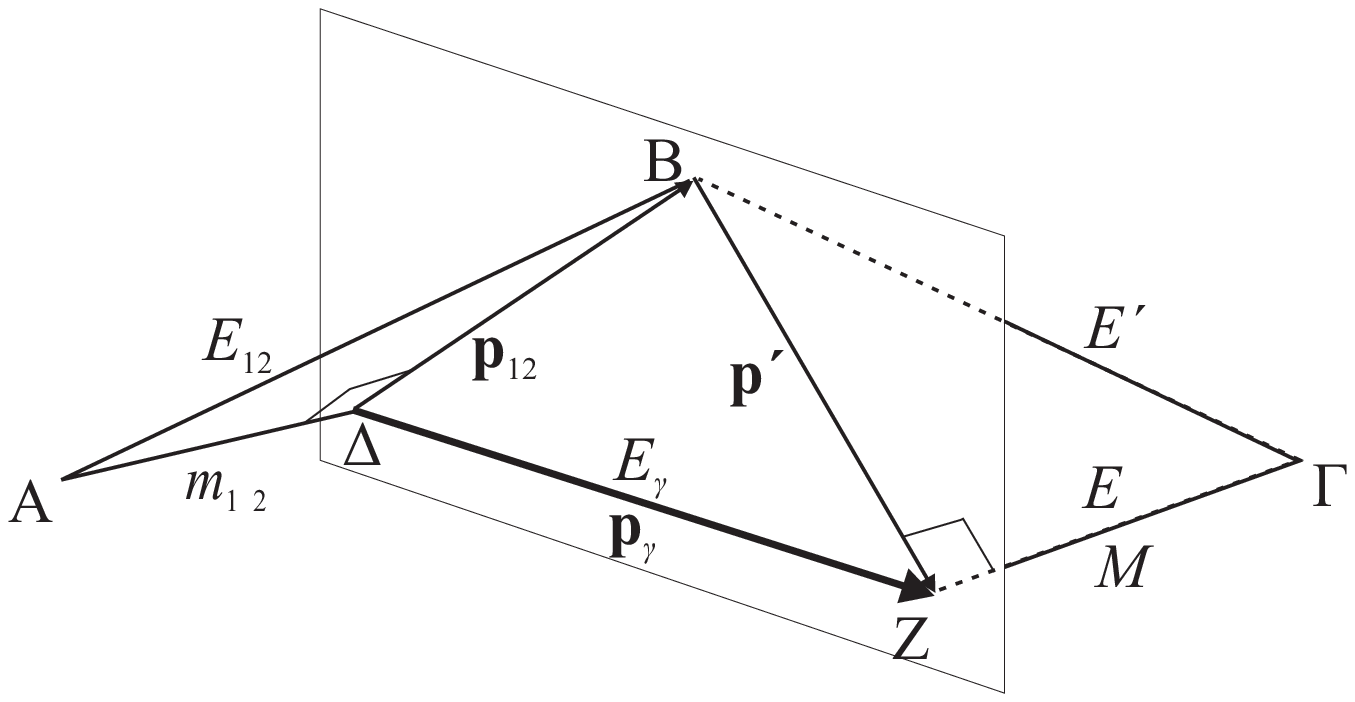}}
\caption{\label{fig:5} The geometric construction corresponding to
the allowed production of a pair of massive particles through
interaction of a photon with a massive target particle.}
\end{center}
\end{figure}
%%%%%%%%%%%%%%%%%%%

Let us depict the photon with a degenerate two-sided triangle ($E_\gamma=|{\bf  p}_\gamma|$),
and the massive particle used to interact with the photon again as a degenerate two-sided triangle (since it is motionless) with
$E=M$ (see Figure \ref{fig:5}). If two new particles ($\#1,\#2$) are finally produced
we draw on momenta screen their total 3-momentum as a ${\bf p}_{12}$ vector and the 3-momentum
of the target particle as ${\bf p}'$. Of course ${\bf p}_{12}+{\bf p}'={\bf p}_\gamma$, while conservation of energy rules that
\be
\textrm{AB}+\textrm{B}\Gamma=E_{12}+E'=E_\gamma+M=\Delta\textrm{Z}+\textrm{Z}\Gamma,
\label{prb2}
\ee
where the subscript $_{12}$ is used to denote the 3-momentum and energy of a composite particle that
corresponds to the two new massive  particles that are produced (see Proposition II) and ${\bf  p}',E'$
are the 3-momentum and the energy of the massive target after the collision, respectively. The composite particle is
depicted by the right triangle $\textrm{A}\Delta\textrm{B}$. The massive target particle, after the reaction will
in general be set in motion, and the corresponding right triangle for this  particle is $\textrm{BZ}\Gamma$.
Therefore the reaction is allowed to take place, as long as the geometric equality of Eq.~(\ref{prb2})
holds good.

%%%%%%%%%%%%%%%%%%%
\begin{figure}[h]
\begin{center}
\scalebox{0.60}{\includegraphics{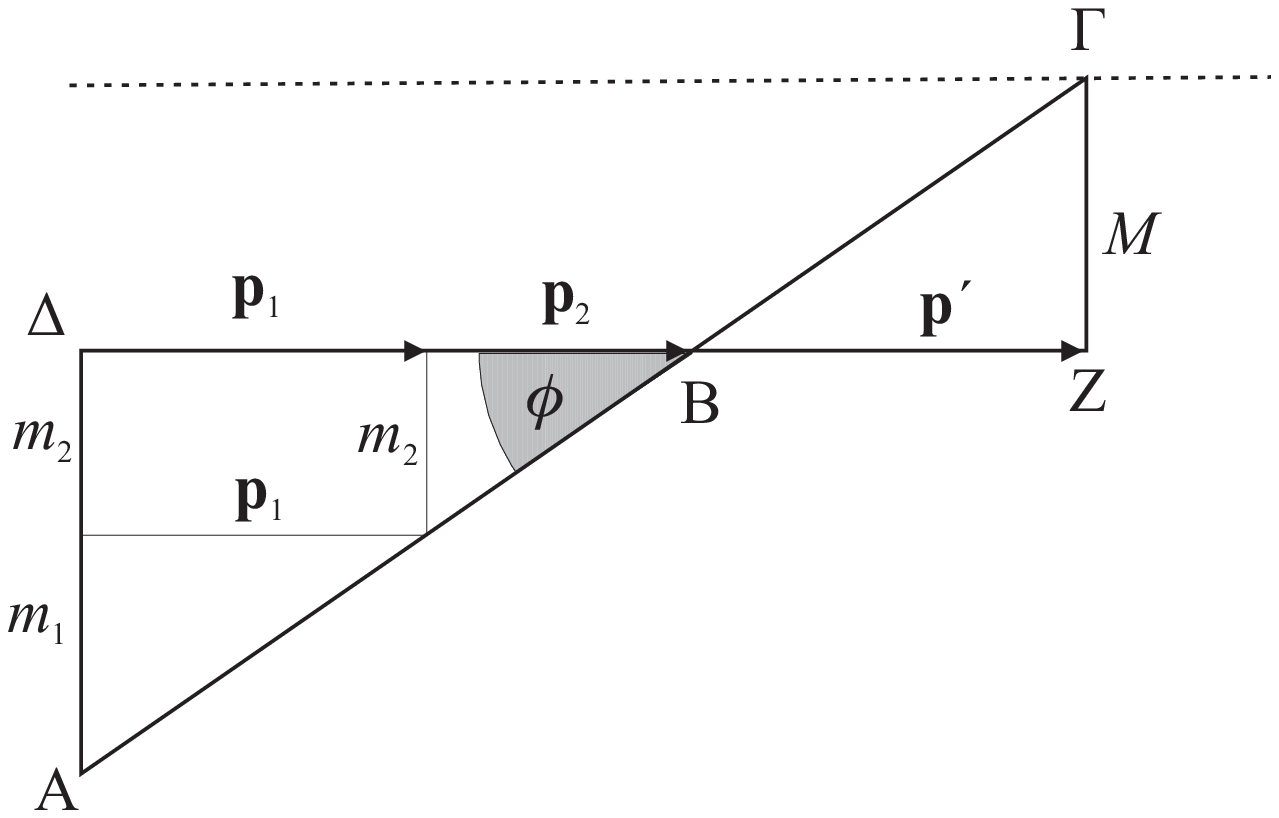}}
\caption{\label{fig:6} The configuration of right triangles which
corresponds to a minimum total energy. This configuration yields the
threshold energy of a photon that could lead to pair production, if
the angle $\phi$ is such that $\Delta \textrm{Z}+\textrm{Z}
\Gamma=\textrm{A} \Gamma$.}
\end{center}
\end{figure}
%%%%%%%%%%%%%%%%%%%

Now let us find the condition for such a reaction to be possible to happen, that is to find
the minimum energy for the photon in order to be able to produce two massive particles.
From Proposition III, the minimum total energy of particles $\#1,\#2$, and the  massive target-particle
is achieved when all 3-momenta ${\bf  p}_1,{\bf  p}_2$, and ${\bf  p}'$ are parallel to each other
and have magnitudes proportional to their masses, $m_{1},m_2$, and $M$, respectively.
This extremal situation, if allowed, sets the threshold condition for the reaction to take place.
Therefore, the necessary condition for the pair production (via a motionless massive target) to take place is the initial
available energy $E_\gamma+M$ to be equal to the minimal total energy of all products. The geometric construction
of triangles that corresponds to this extremal situation is shown in Figure \ref{fig:6}, and the
required condition is to find a point $\Gamma$ along the dashed line which lies at a distance $m_1+m_2+M$
from point $\textrm{A}$ so that the length of $\textrm{A}\Gamma$ equals the length of $\Delta\textrm{Z}+\textrm{Z}\Gamma$.
It is intuitively obvious that such a point $\Gamma$ always exists, since while the point $\Gamma$ moves along the dashed line
the difference between the two lengths varies continuously, with
the former length ($\textrm{A}\Gamma$) being longer than the latter one ($\Delta\textrm{Z}+\textrm{Z}\Gamma$) when
the point $\Gamma$ lies near the projection of $\textrm{A}\Delta$ on the dashed line, while the two lengths are in
opposite relation to each other when  point $\Gamma$ is moved far away ($\Delta\textrm{Z}>>m_1+m_2+M$) since then
\begin{widetext}
\be
\textrm{A}\Gamma=\sqrt{(\Delta\textrm{Z})^2+(m_1+m_2+M)^2}
\simeq \Delta\textrm{Z}
<\Delta\textrm{Z}+M.
\ee
Therefore, whatever the mass of the target, there is always a sufficiently energetic  photon
that could lead to the production of the two specific massive particles. Let us now quantify this conclusion. From the
triangles depicted in Figure \ref{fig:6} we have
\be
(E_1+E_2+E')_{\min}=\textrm{A}\Gamma=E_{\gamma}^{\textrm{(thres)}} +M=\textrm{A}\Gamma(\cos\phi+\frac{M}{M+m_1+m_2}\sin\phi),
\ee
where $\phi=\widehat{\Delta\textrm{BA}}$.
This leads to the algebraic relation
\be
\cos\phi+\frac{M}{M+m_1+m_2}\sin\phi=1.
\ee
Expressing all trigonometric functions shown up in the above formula in terms of $\tan\phi$, we obtain the solution for $\phi$
\be
\tan\phi=\frac{2 M (M+m_1+m_2)}{(M+m_1+m_2)^2-M^2}.
\ee
Therefore the photon should have at least energy equal to
\be
E_\gamma^{\textrm{(thres)}}=
|{\bf  p}_\gamma^{\textrm{(thres)}}|=
\frac{M+m_1+m_2}{\tan\phi}=(m_1+m_2)\left(1+\frac{m_1+m_2}{2 M} \right),
\label{Ethresh2}
\ee
in order to be able to produce the particular pair of particles.
\end{widetext}

If the photon has energy higher than this
minimum value, the conservation of energy is not satisfied by this optimum configuration
(it would then be  $\Delta\textrm{Z}+\textrm{Z}\Gamma>\textrm{A}\Gamma$ in Fig.~\ref{fig:6}).
One should then arrange the 3-momenta of all particles after the reaction
in such a way  that the total energy exceeds this minimum total energy by breaking the constraint of having
all 3-momenta corresponding to equal velocities for all three particles, thus permitting the line $\textrm{A}\Gamma$
to be a broken one as in Fig.~\ref{fig:5}.

\begin{figure}[h]
\begin{center}
\scalebox{0.60}{\includegraphics{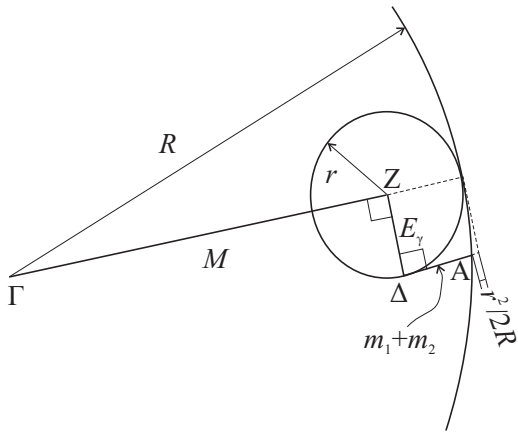}}
\protect\caption{\label{fig:7} In the case of a very massive target,
the configuration that corresponds to the threshold for a  pair
production is represented by the depicted geometrical structure:
$\Gamma\textrm{Z}=M$, $\textrm{Z}
\Delta=r=E_\gamma^{\textrm{(thres)}}=|{\bf
p}_\gamma^{\textrm{(thres)}}|$,
$\Delta \textrm{A}=m_1+m_2$, and $\Gamma \textrm{A}=R=$ %\linebreak[4]
$M+E_\gamma^{\textrm{(thres)}}$.
From this drawing it is obvious that for $R>>r$, $m_1+m_2=\Delta \textrm{A} \protect\cong r=
E_\gamma^{\textrm{(thres)}}$. The error of this approximation is of order $r^2/2R$ (deviation of a circle
from its tangent).}
\end{center}
\end{figure}

From Eq.~(\ref{Ethresh2}) we observe that
for $M>>m_1+m_2$, the threshold energy for the photon is approximately
$E_\gamma^{\textrm{(thres)}} \simeq m_1+m_2$. Now we will directly demonstrate this fact geometrically without
going through  the general result (\ref{Ethresh2}).
We draw a segment of length $\Gamma\textrm{Z}= M$,
and with center $\textrm{Z}$ we draw a circle of radius $r=E_\gamma<M$.
We also draw another circle of radius $R=E_\gamma+M$ (which is the initial energy of the system),
having its center at $\Gamma$. The broken line $\Gamma \textrm{Z} \Delta \textrm{A}$ (see Fig.~\ref{fig:7})
with $\widehat{\textrm{Z}}=\widehat{\Delta}=90^\circ$, and $\Delta \textrm{A}=m_1+m_2$ constitutes
the configuration of masses and 3-momenta for the products of the reaction which corresponds to the minimum
total energy (cf.~Fig.~\ref{fig:6}).
In this case the initial 3-momentum of the photon is distributed proportionally to $M$ and $m_1,m_2$ after the reaction in such a way that
all particle-triangles are similar to each other.
From the drawing in Fig.~\ref{fig:7} it is clear that the radius of the small circle $E_\gamma^{\textrm{(thres)}}$
should be approximately equal to $\Delta \textrm{A}$, that is equal to $m_1+m_2$ when $M>>m_1+m_2$ with a relative error  of
order $r/(2 R)=(m_1+m_2)/[2(M+m_1+m_2)]  \cong (m_1+m_2)/(2 M)$. This error represents the missing segment in order to form an exact
square that is shown in Fig.~\ref{fig:7}, and it actually leads to the exact expression for the threshold energy (see Eq.~\ref{Ethresh2}).

%%%%%%%%%%%%%%%%%%%
%%%%%%%%%%%%%%%%%%%

%%%%%%%%%%%%%%%%%%%%%%%%%%%%%%%%%%%%%%%%%%%%%%
\section{Problem 3}
\label{sec:5}
%%%%%%%%%%%%%%%%%%%%%%%%%%%%%%%%%%%%%%%%%%%%%%

The following problem is something like a generalization of Problem 2. Let us assume that
a particle ${\cal A}$ hits a motionless target-particle ${\cal B}$, and   a number of particles ${\cal C}_i$ ($i=1,2,\ldots$)
emerge after the collision (the initial particles could be among the products). The question is what is
the minimum energy of particle ${\cal A}$ in order to render this particular reaction possible to happen.
Any particle, apart of ${\cal B}$, could be massless.

Following the same line of arguments as in Problem 2, we look for the minimum energy required to
obtain the optimal final configuration, which is the one with all particles moving
with the same velocity (according to proposition III). We now discern two cases:
(a) The produced particles have total mass less than, or equal to,
the total mass of the initial particles.
(b) The produced particles have  larger total mass than the initial particles.

\begin{figure}[h]
\begin{center}
\scalebox{0.7}{\includegraphics{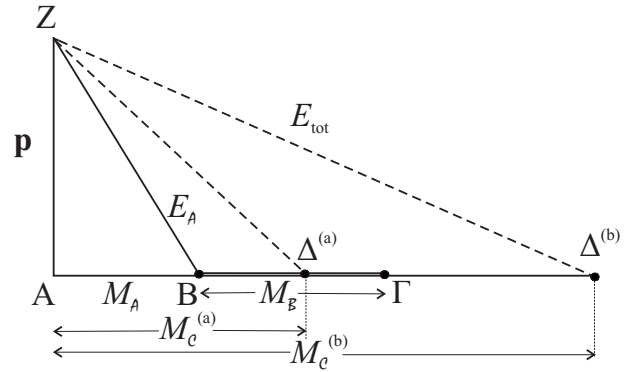}}
\caption{\label{fig:8} Both cases, (a) and (b), are depicted in this
diagram. Case (a): when $M_{\cal C} \leq M_{\cal A}+M_{\cal B}$, in
order to comply with energy conservation, we should have a non
optimum configuration; therefore there is no restriction for
$E_{\cal A}$. Case (b): when $M_{\cal C} > M_{\cal A}+M_{\cal B}$,
there is a minimum value of $|{\bf  p}_{\cal A}|$, and thus a
minimum value of $E_{\cal A}$, for which the reaction is allowed to
happen by following the optimal energy configuration for the
products.}
\end{center}
\end{figure}

In case (a) the optimum configuration cannot comply with the conservation of energy.
The argument goes as follows: if the total mass of the products was equal to the total
mass of the initial particles, the optimal configuration for the products would correspond
to a right triangle (see Fig.~\ref{fig:8}) with vertical sides equal to
the total 3-momentum ($\textrm{ZA}$), and the total mass of them ($\textrm{A}\Gamma$), respectively (remember
that in the minimum energy configuration --according to Proposition III--
all particles are moving at the same speed; consequently
all right triangles are similar and they could all be combined in a single right triangle with its orthogonal
sides equal to the sum of all masses,
and the total 3-momenta, respectively, while the hypotenuse is equal to the total energy.)
Thus the energy would be
represented by the segment $\textrm{Z}\Gamma$ .
However, from triangular inequality, this is shorter than $\textrm{ZB}+\textrm{B}\Gamma$
which represents the total energy of the initial particles. For lower total mass of products, the inequality is even stronger;
then the segment $\textrm{Z}\Delta^{(a)}$
in Figure \ref{fig:8}, which represents the energy of the produced particles
in their optimum configuration is shorter than the available energy of the initial particles.
Thus in case (a) we could increase the final energy (in order to comply with energy conservation),
by choosing a non optimal configuration for the products, without changing the total
3-momentum of the system of particles. Thus the reaction could happen by
following such a non-optimal configuration of 3-momenta.

In case (b) we will look for the minimum required 3-momentum of particle ${\cal A}$ to make conservation of energy hold under the
optimum configuration of the products; if $|{\bf  p}_{\cal A}|=|{\bf  p}|$ exceeds this threshold value,  a non-optimal arrangement
of products' 3-momenta could again be followed to make conservation of energy hold.

%%%%%%%%%%%%%%%%%%%
%%%%%%%%%%%%%%%%%%%

In Figure \ref{fig:8} we have also drawn the triangle that corresponds to the optimal configuration for the products
in case (b), which is supposedly compatible with the energy conservation:
$\textrm{AZ}$ is the threshold magnitude of 3-momentum of particle ${\cal A}$ (and
thus of all products), $\textrm{AB}$ is the mass of the incident particle, ${\cal A}$, $\textrm{B}\Gamma$
is the mass of the target particle ${\cal B}$, and $\textrm{A}\Delta^{(b)}=M_{\cal C}$ is the sum of masses of all products.
From conservation of energy $E_\textrm{tot}=E_{\cal A}+M_{\cal B}$. Thus
\be
\begin{split}
(\textrm{AZ})^2 + (\textrm{A}\Delta^{(b)})^2= &|{\bf p}|^2 + (M_{\cal C}^{(b)})^2 = \\
E_\textrm{tot}^2   =& (E_{\cal A}+M_{\cal B})^2,
\end{split}
\ee
and
\be
(\textrm{AZ})^2 + (\textrm{AB})^2           = & |{\bf p}|^2 +  (M_{\cal A})^2
= E_{\cal A}^2.
\ee
Subtracting these two relations and solving for $E_{\cal A}$ we obtain the
desired threshold energy
\be
E_{\cal A}=\frac{M_{\cal C}^2 - M_{\cal A}^2 - M_{\cal B}^2 }{2 M_{\cal B}},
\ee
which is definitely positive since $M_{\cal C}^{(b)} > M_{\cal A} + M_{\cal B}$.
We should note once again that this final result generalizes the result of the previous problem since
by setting $M_{\cal A}=0$ (photon), $M_{\cal B}=M$,  and $M_{\cal C}^{(b)}=m_1+m_2+M$ we obtain Eq.~(\ref{Ethresh2}).

%%%%%%%%%%%%%%%%%%%%%%%%%%%%%%%%%%%%%%%%%%%%%%
\section{Problem 4}
\label{sec:7}
%%%%%%%%%%%%%%%%%%%%%%%%%%%%%%%%%%%%%%%%%%%%%%

What are the possible arrangements of the 3-momenta of two particles
as a result of an elastic collision between them? By elastic collision we mean that
the two particles remain unaltered after the collision, preserving their
total kinetic energy, while no new particles are created.

This is a
problem that most clearly demonstrates the power of geometric
constructions in getting qualitative results with respect to relativistic
particle collisions.
The classical analytic solution (see \cite{LandauRela}) requires first to shift to
the center-of-momentum frame by applying Lorentz transformations on the initial
4-momenta, analyze  the possible arrangements of the particles' 4-momenta in this particular frame of reference
after the collision, and finally go back to laboratory frame by performing
an inverse Lorentz transformation of the post-collision 4-momenta.

In our graphical method, we draw
on momenta screen a  hypothetical possible configuration of the two particles' 3-momenta
after collision. From conservation of total 3-momentum these vectors
should sum up to the total 3-momentum of the particles before collision. Thus
the two vectors on momenta-screen should be the two sides of a triangle with a fixed
third side (the total 3-momentum).
On the other hand, conservation of total energy
means that the two 3-momentum vectors should have such magnitudes that the
corresponding energies sum up to a fixed value (the total  energy of the initial particles).
Actually, the specific arrangement of the particles'
3-momenta before collision is one of the solutions of all possible configurations
we are looking for.
One should keep in mind that the orientation of the plane of momenta screen
is not known a priori, since only the total 3-momentum vector is given;
thus the plane on which all configurations of 3-momenta
that are compatible with conservation of both 3-momentum and energy are drawn,
should be rotated around the line of total 3-momentum to get all possible configurations of 3-momenta in space.
For the moment we will just fix the orientation of this plane to find the locus of the heads of
the particles' 3-momentum vectors that correspond to all possible
relative arrangements of the two particles' 3-momenta on this particular plane.

%%%%%%%%%%%%%%%%%%%
\begin{figure}[h]
\begin{center}
%\scalebox{0.50}{\includegraphics{fig10new.eps}}
\centerline{\includegraphics[width=18pc] {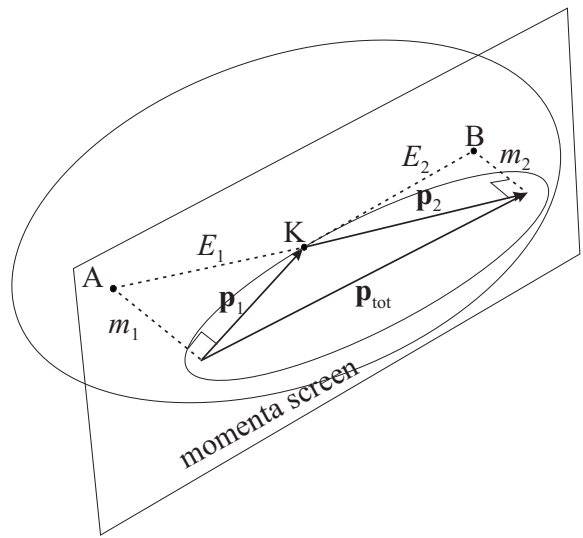}}
\caption{\label{fig:10} The intersection of the ellipsoid of
constant energy by the plane of momenta screen provides all possible
configurations of particles' 3-momenta in an elastic collision.}
\end{center}
\end{figure}
%%%%%%%%%%%%%%%%%%%

To find these arrangements, we
draw two segments (see Fig.~\ref{fig:10}) of length $m_1,m_2$,
respectively, perpendicular to this plane so that
one edge of the first segment is the tail of the first particle's
3-momentum (which is also the tail of the total 3-momentum),
while one edge of the second segment is the head of the second particle's
3-momentum (which is also the head of the total 3-momentum).
Both segments have been drawn
on the same side of the plane (although they could be drawn on opposite sides
without affecting the results).
Then, the broken line AKB connecting the free edges of
the perpendicular mass-segments, A, B, which lie out of the plane, through point K which lies on moment screen (this
point is representing
a possible arrangement of the two 3-momenta vectors) has length equal
to the total energy $E_1+E_2$ of the two particles (AK and KB are the hypotenuses
of the two right triangles that correspond to
the two particles).  The latter quantity
is fixed and equals the total energy of the particles before collision.
Thus all possible arrangements of 3-momenta are described by the locus of a point
that lies on momenta screen while the sum of  its distances from the given points A, B
is equal to the total energy. Obviously the  points K that satisfy the above requirements are described by
the intersection of the plane of momenta-screen with the axially symmetric
ellipsoid with focuses A, B and major axis equal to the total energy of the particles.
The intersection of a plane with an ellipsoid is definitely an ellipse.
The explanation is simple. Such an intersection is clearly a closed curve, and since
the ellipsoid is described by a quadratic polynomial of cartesian coordinates, its
intersection by a plane is described by a quadratic relation as well.
However, the most general closed plane curve of quadratic form is an ellipse. As mentioned above, we should
finally rotate this elliptical curve around the line of the
total 3-momentum to obtain all possible configurations of ${\bf p}_1,{\bf p}_2$.
It is easy to see that this line coincides with the major axis of the
constructed ellipse, by reflection symmetry of the whole 3-D construction with respect to
the plane defined by the two parallel mass segments (see Fig.~\ref{fig:10}).

It is remarkable that we have arrived at this conclusion,
with respect to all possible
arrangements of 3-momenta, by simple geometric arguments without resorting to any
mathematical relations at all.

%%%%%%%%%%%%%%%%%%%
\begin{figure}[h]
\begin{center}
%\scalebox{0.50}{\includegraphics{fig10newnew.eps}}
\centerline{\includegraphics[width=20pc] {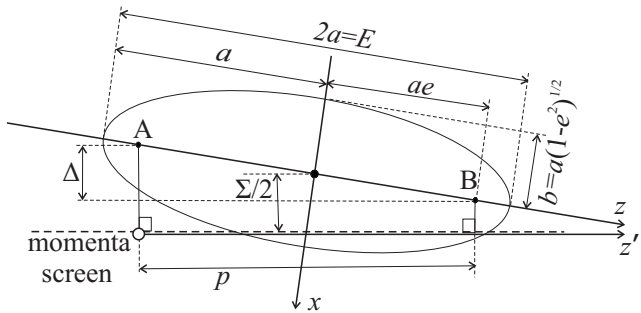}}
\caption{\label{fig:11} A 2-D projection (a ground plan) of
Fig.~\protect\ref{fig:10} on which the various lengths related to
the ellipsoid are written down. The semi-major axis of the ellipsoid
is $a=E/2$, while the focal distance $\textrm{AB}$ is $2 a
e=\sqrt{p^2+\Delta^2}$ ($e$ is the eccentricity of the ellipsoid).
Finally the semi-minor axis is $b=a
\sqrt{1-e^2}=(\sqrt{E^2-p^2-\Delta^2})/2$. All these parameters have
been used to construct the equation describing the plane of momenta
screen which is drawn here as a line that intersects both $z$- and
$x$-axis. The $y$-axis is perpendicular to the plane of the
diagram.}
\end{center}
\end{figure}
%%%%%%%%%%%%%%%%%%%

Now let us proceed to quantify our geometric result.
We will use the following notation to simplify our formulae:
$E \equiv E_1+E_2$ is the total energy of particles, $p \equiv |{\bf  p}_\textrm{tot}|=|{\bf  p}_1+
{\bf  p}_2|$ is the magnitude of the total 3-momentum,
$\Delta \equiv |m_1-m_2|$ and $\Sigma \equiv m_1+m_2$.
First we write the equation that describes the ellipsoid with focuses
A, B, in cartesian coordinates:
\be
\frac{z^2}{E^2}+\frac{x^2+y^2}{E^2-p^2-\Delta^2}=\frac{1}{4},
\label{ellipseq}
\ee
where the $z$-axis lies along the line  $\textrm{AB}$  while the $x$- and
$y$-axes are perpendicular to the $z$-axis,
with the $y$-axis being parallel to the plane of momenta screen.
The origin of this cartesian coordinate system is at the
middle point of the segment $\textrm{AB}$, which is also the center of the ellipsoid.
On the other hand the plane of momenta screen in the same coordinate system is described by the equation:
\be
x=-\frac{\Delta}{p} \left(z- \frac{\Sigma \sqrt{p^2+\Delta^2}}{2 \Delta} \right),
\label{planeeq}
\ee
as one could infer from the lengths of various segments shown in Fig.~\ref{fig:11}.
Finally if we use a  new $z'$-axis defined as the line formed by the intersection of momenta screen by the plane $y=0$,
with its origin ($z'=0$) coinciding with the origin of ${\bf  p}_\textrm{tot}$ (empty circle on the
$z'$ axis in Fig.~\ref{fig:11}),
the relation between the old $z$- and the new $z'$-values of a point
along the $z'$-axis is
\be
z'=z \frac{\sqrt{p^2+\Delta^2}}{p}+\frac{p^2-\Sigma \Delta}{2 p}.
\ee
Now, if we seek for a simultaneous solution of Eq.~(\ref{ellipseq})
and of Eq.~(\ref{planeeq}) (intersection of the ellipsoid by the plane of momenta screen)
and we replace the $z$-values
by its equivalent $z'$-values we arrive at the general form
of the ellipse describing the locus of the head points
of all possible configurations of 3-momenta vectors
after the collision. The corresponding equation is
\begin{widetext}
\be
\left( 1-\frac{p^2}{E^2} \right)
\left[ p_{\parallel} - \frac{p}{2} \left(1+ \frac{\Sigma \Delta}{E^2-p^2} \right) \right]^2
+ p_{\perp}^2=\frac{(E^2-p^2-\Delta^2)(E^2-p^2-\Sigma^2)}{4(E^2- p^2)},
\label{theellipse}
\ee
\end{widetext}
which is clearly the equation of an ellipse, as it was anticipated by the aforementioned
geometric arguments. Since on the plane of  momenta screen
we draw the 3-momenta vectors of the particles after the collision,
we have replaced the $z'$ and the $y$, with $p_{\parallel}$
and $p_{\perp}$, respectively.
The $_{\perp}$ and $_{\parallel}$ notation corresponds to the
components of each 3-momentum vector that is parallel or perpendicular to the
total 3-momentum vector. Constructing the equation of the ellipse that is describing all possible
configurations of 3-momenta might be a bit tedious but it is straightforward compared to
the usual analytic method.

We could also draw further conclusions from our geometric construction alone,
without any reference to Eq.~(\ref{theellipse}).
\noindent (a) In the non-relativistic limit
the two $m_i$-segments ($i=1,2)$ are so large
with respect to the magnitude of the total 3-momentum ($m_i >> |{\bf p}_i|$), that
we could think of both mass segments as lying along the same line perpendicular to the plane
of momenta screen.
In this case the ellipsoid of constant energy is an extremely elongated ellipsoid
oriented so that its axis of symmetry (major axis) is almost perpendicular to the plane
of momenta screen.
The intersection of such an ellipsoid with the plane
of momenta is the well known circular locus of momenta vectors in elastic
collision in classical mechanics (see \cite{LandauMech}).

%%%%%%%%%%%%%%%%%%%
\begin{figure}[h]
\begin{center}
%\scalebox{0.50}{\includegraphics{fig10newnew.eps}}
\centerline{\includegraphics[width=18pc] {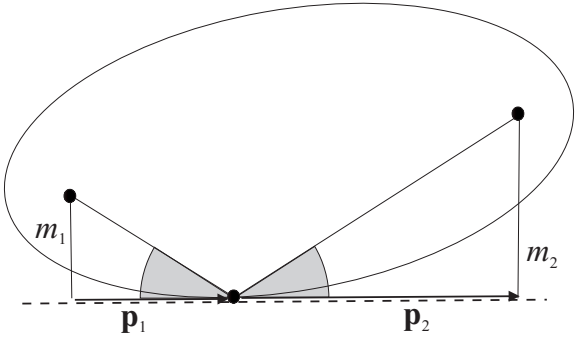}}
\caption{\label{fig:12} When the moment screen (dashed line) is
tangent to the ellipsoid of constant energy there is only a single
arrangement of 3-momenta. The two equal angles correspond to equal
velocities for the two particles; thus corresponds to a
non-collisional configuration (case b).}
\end{center}
\end{figure}
%%%%%%%%%%%%%%%%%%%

\noindent
(b) In the relativistic regime the extreme case of a single point intersection
of the ellipsoid with the momenta screen  corresponds to a configuration where
the plane of momenta is tangent to the ellipsoid. Due to reflection symmetry with
respect to the $x-z$ plane (cf.~Fig.~\ref{fig:10}), this intersecting point is along
the direction of the total 3-momentum vector (see Fig.~\ref{fig:12}); therefore
the two particles can only move along the same direction.
Moreover, there is a geometric property according to which the tangent line to an ellipse
forms equal angles with the focal radii at the point of contact \cite{AnalGeom}.
The tangents of these angles in our geometric construction are nothing but the two particles' velocities; thus this
singular case corresponds to two particles moving with exactly the same speed, one behind the other.
Of course such particles will never collide;  they will always move maintaining
their distance. So a single point intersection in our geometrical solution corresponds to this trivial
physical configuration.

%%%%%%%%%%%%%%%%%%%
\begin{figure}[h]
\begin{center}
%\scalebox{0.50}{\includegraphics{fig10newnew.eps}}
\centerline{\includegraphics[width=18pc] {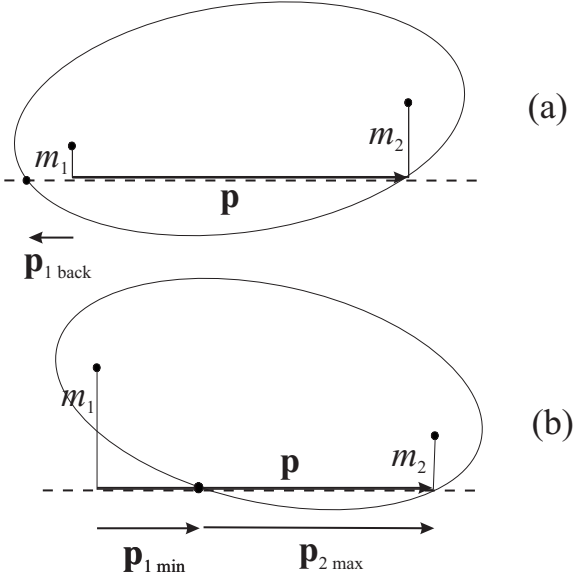}}
\caption{\label{fig:13} The relative position of the ellipsoid and
the plane of momenta screen corresponding to the two cases of
comment (c) where ${\bf p}_2=0$: The top drawing (a) is for
$m_1<m_2$ and the bottom (b) for $m_1>m_2$. Both drawings are planar
cross sections perpendicular to the momenta-screen (dashed line) as
in Figs~\protect\ref{fig:11},\protect\ref{fig:12}.}
\end{center}
\end{figure}
%%%%%%%%%%%%%%%%%%%

\noindent
(c) When one of the particles --let's say  particle $\#2$--
is initially at rest, the ellipsoid intersects
the momenta screen at  one of the edges of the vector ${\bf p}$. This happens
because the initial (before the collision) configuration
$({\bf  p}_2=0, {\bf  p}_1={\bf  p})$  should correspond to one of all possible
arrangements of 3-momenta vectors. In this case, that point
is the terminal point of ${\bf  p}$. Now it is clear that if particle $\#1$
is the lighter one (see Fig.~\ref{fig:13}(a)) the elliptical locus of the possible
arrangements of ${\bf  p}_1$ (this ellipsis is on momenta screen, therefore it is not shown in Fig.~\ref{fig:13})
intersects the line of ${\bf  p}$ at the terminal point
of ${\bf  p}$ and at a second point that lies along the opposite direction of ${\bf  p}$.
This means that the light particle could be backscattered.
On the other hand for a heavy particle $\#1$ (with respect to particle $\#2$) the corresponding situation is depicted in
Fig.~\ref{fig:13}(b). The second intersecting point of the ellipsoid
with the plane of 3-momenta now
is an intermediate point along the vector ${\bf  p}$. Therefore,
a heavy incident particle cannot be backscattered. Moreover the maximum angle
at which particle $\#1$ could be deflected corresponds to the tangent line from
the starting point of ${\bf  p}$ to the ellipse on momenta screen that
describes all 3-momenta arrangements. Algebraic manipulations of the standard form
of the equation for the tangent line of an ellipse from a given point \cite{AnalGeom} lead to
the following value for this angle:
\be
\tan\theta_{1, \max}=\frac{P_\perp}{\sqrt{(P_1+P_2/2)^2-(P_2/2)^2}},
\label{Ps}
\ee
%%%%%%%%%%%%%%%%%%%
\begin{figure}[h]
\begin{center}
%\scalebox{0.50}{\includegraphics{fig10newnew.eps}}
\centerline{\includegraphics[width=18pc] {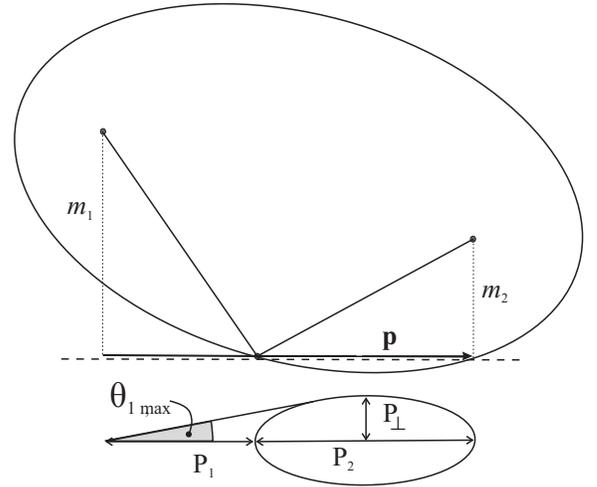}}
\caption{\label{fig:14} Top: A ground plan of the ellipsoid of
constant total energy with respect to momenta screen (dashed line).
Bottom: The elliptical locus of all possible arrangements for the
two 3-momenta which arises from the intersection of the ellipsoid
with the plane of momenta screen. This is the portrait of momenta
screen corresponding to the above relative position of the ellipsoid
and the momenta screen. The $P_\perp, P_1$ and $P_2$ lengths used in
Eq.~\protect{\ref{Ps}} are shown on this diagram.}
\end{center}
\end{figure}
%%%%%%%%%%%%%%%%%%%
where $P_1$ is the minimum magnitude of ${\bf  p}_1$ (it is equal to that part of ${\bf  p}$
that is left out of the ellipsoid),  $P_2$ is the  maximum magnitude of
${\bf  p}_2$ (it is the part of ${\bf  p}$ that lies inside the ellipsoid),
and $P_\perp$ is the semi-minor axis of the ellipse (the maximum magnitude of ${\bf p}_1$
and ${\bf p}_2$ that is perpendicular to ${\bf  p}$). All these elements
are easily computed from expression (\ref{theellipse}). The corresponding expressions
yield the following values:
\begin{widetext}
\be
\begin{split}
P_\perp=\max [p_\perp ]=\sqrt{ \frac{(E^2-p^2-\Delta^2)(E^2-p^2-\Sigma^2)}{4(E^2- p^2)} } \\%-
% \left(1-\frac{p^2}{E^2}\right)
% \frac{p^2}{4} \left(1+ \frac{\Sigma \Delta}{E^2-p^2} \right)^2 } \\
P_1=\min [ p_{\parallel}|_{p_\perp=0}]=\frac{p}{2} \left(1+ \frac{\Sigma \Delta}{E^2-p^2} \right)
-\sqrt{ \frac{(E^2-p^2-\Delta^2)(E^2-p^2-\Sigma^2) E^2}{4(E^2- p^2)^2} }\\
P_2=\max [p_{\parallel}|_{p_\perp=0}] - \min [p_{\parallel}|_{p_\perp=0}]=
\sqrt{ \frac{(E^2-p^2-\Delta^2)(E^2-p^2-\Sigma^2) E^2}{(E^2- p^2)^2} }.
\end{split}
\ee
After quite some algebra we obtain the following simple value for this angle:
\be
\tan\theta_{1, \max}=\frac{m_2}{\sqrt{m_1^2-m_2^2}}.
\ee
Oddly enough, this happens to be exactly the answer one gets for the analogous
classical non-relativistic elastic collision (see \cite{LandauMech}).
\end{widetext}

\begin{figure}[h]
\begin{center}
%\scalebox{0.50}{\includegraphics{fig10newnew.eps}}
\centerline{\includegraphics[width=18pc] {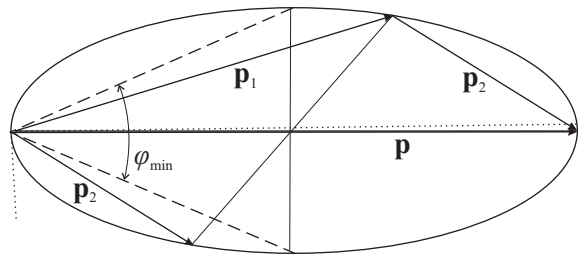}}%
\caption{\label{fig:15} The 3-momenta $\protect{\bf
p}_1,\protect{\bf p}_2$ of two billiard balls after an elastic
collision are represented by two vectors that sum up to the major
axis of an ellipse, the total 3-momentum $\protect{\bf p}$. The
angle between them is the inscribed angle from the leftmost point
of the ellipse that subtends one of the diameters of the ellipse.
This  angle ranges from  $90^\circ$ (approximately the angle formed
by the dotted chords), to $\phi_{\min}$ (dashed chords).}
\end{center}
\end{figure}

\noindent
(d) In a relativistic billiard game (the analogue of the classical non-relativistic one), the major axis of the
ellipse describing all 3-momenta configurations is equal to the magnitude of the total 3-momentum (one of the two identical
balls is assumed to be initially at rest). Hence the angle
between the lines of motion of the two balls after their elastic collision  is the inscribed angle
which subtends a diameter of the elliptical locus of 3-momenta (cf.~Figure \ref{fig:15})
and with its vertex at the starting point of the total
${\bf  p}$ (the leftmost point of the ellipse).
This angle lies between $90^\circ$ and $\phi_{\min} =2 \tan^{-1}(b/a)$, where $b,a$ are the semi-minor
and the semi-major axis of the ellipse, respectively. The former happens when one of the balls
is almost still after the collision (the angle formed by the dotted lines in Fig.~\ref{fig:15}),
while the latter corresponds to the case of equal 3-momenta after collision.
The non-relativistic locus of momenta is a circle instead of an ellipse (cf.~case (a)); all inscribed
angles then that subtend to a diameter are equal to $90^\circ$, and the balls always move perpendicular     to each other
after the collision, as is well known by billiard players.

%%%%%%%%%%%%%%%%%%%
%%%%%%%%%%%%%%%%%%%

%%%%%%%%%%%%%%%%%%%%%%%%%%%%%%%%%%%%%%%%%%%%%%
\section{Problem 5}
\label{sec:8}
%%%%%%%%%%%%%%%%%%%%%%%%%%%%%%%%%%%%%%%%%%%%%%

In beta decay, what is the range of kinetic energy of the electron
produced, and what is the distribution function of that energy?

This is a reaction where a single particle (in the case of beta decay this particle is a neutron)
spontaneously disintegrate into three particles;
one of them (an electron antineutrino) being a very light one.
It was first found by L. Meitner and O. Hahn that the electrons produced in beta decay have a
continuous spectrum although only two particles (an electron and a proton) were apparently produced.
As we will demonstrate through geometrical arguments, production of
only two particles is not compatible with a continuous spectrum.
Pauli suggested that production of a third light particle,
which could not be detected then, could resolve the mystery.
Fermi baptized that particle neutrino, and it was actually detected
a quarter of a century later.

%%%%%%%%%%%%%%%%%%%
\begin{figure}[h]
\begin{center}
\scalebox{0.60}{\includegraphics{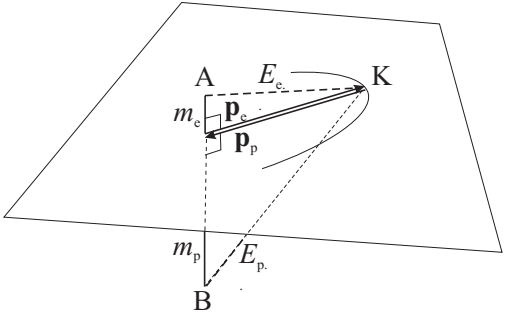}}
\caption{\label{fig:16} There is only one way to distribute the
total energy in  two particles that are produced from the
spontaneous decay of a neutron. The two particles should have
opposite 3-momenta with the same magnitude. These magnitudes should
be compatible with the aforementioned conservation of energy. The
mutual line of motion, though, could have any direction.}
\end{center}
\end{figure}
%%%%%%%%%%%%%%%%%%%

Let's first assume that a neutron, which is initially at rest, spontaneously decays into two particles; a proton and an electron.
On momenta screen we  draw (cf.~Figure \ref{fig:16}) the two 3-momenta vectors so that they add  up to zero
(since the parent particle is assumed to be still
in the lab frame). Two segments of length $m_p$ and $m_e$ are drawn perpendicular to the momenta-screen plane at one of the
common edges of the two opposite 3-momenta, on each side of the plane.
The length of the broken line AKB connecting the second common edges of 3-momenta, K, with the free
edges A, B of the mass segments ($m_e$ and $m_p$ respectively),
equals the total energy of the system, that is the rest mass $m_n$ of the neutron.
It is intuitively obvious that there is only one solution for the magnitude of 3-momenta
of the daughter particles (apart from their direction).
This solution is represented by the radius of the circular intersection of an ellipsoid
--with its foci at A, B, and with major axis equal to $m_n$--
with the plane of momenta screen.  Therefore, if there was no extra particle produced,
the electron should be monoenergetic.

Now if we allow for one more particle (more than one extra particles could be considered equivalent to a single
composite particle; cf.~Proposition II),
there is a continuous sequence of arrangements for the 3-momenta of proton and electron, and correspondingly energies,
that could accommodate an extra particle. More specifically we will
focus our attention on a zero-rest-mass particle, like what was assumed for many years for the neutrinos (practically it
could be considered as such, compared to the other two heavy-mass particles).
The three 3-momenta drawn on momenta screen should still sum up to zero, forming a generic triangle $\textrm{K}\Lambda \textrm{M}$.
Since the new particle has no rest mass, the three-segment broken line $\textrm{A} \textrm{K}  \Lambda \textrm{B}$
($\textrm{K}\Lambda$ representing the new particle's 3-momentum),
should have total length equal to $m_n$. It is easy to see that we
could continuously deform the triangle of 3-momenta $\textrm{K}\Lambda \textrm{M}$
while keeping the total energy (represented by the length of $\textrm{A} \textrm{K}  \Lambda \textrm{B}$) fixed.
This is a simple pictorial  argument
that demonstrates why the  electrons in beta decay should come out with a continuous spectrum.

%%%%%%%%%%%%%%%%%%%
\begin{figure}[h]
\begin{center}
\scalebox{0.70}{\includegraphics{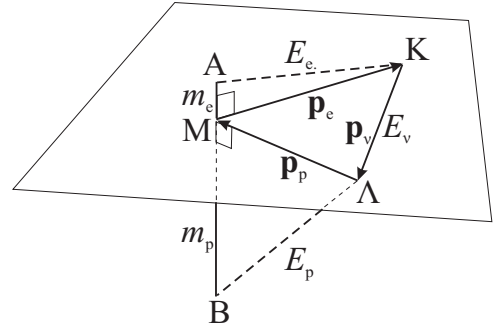}}
\caption{\label{fig:17} A third massless particle among the products
allows for a wide range of energies for the electron. Any choice for
the points $\textrm{K}$, $\Lambda$ on momenta screen, such that
$\textrm{AK}+\textrm{K}\Lambda+\Lambda\textrm{B}=m_n$, represents a
distinct configuration of 3-momenta for the products of beta decay.
}
\end{center}
\end{figure}
%%%%%%%%%%%%%%%%%%%

Next we will turn into a more quantitative analysis of all possible arrangements for the energies of the three particles.
First of all we can convince ourselves that an electron could be produced with no kinetic energy at all. This arises when the
$\textrm{KM}$ side of the triangle of 3-momenta has zero length ($\textrm{K}$ and $\textrm{M}$ points coincide),
while the third vertex is such that
\be
\textrm{AM}+\textrm{M}\Lambda+\Lambda\textrm{B} &=& m_e+\textrm{M}\Lambda+\sqrt{(\textrm{M}\Lambda)^2+m_p^2} \nn \\
&=& m_n.
\ee
This situation is unique with respect to the magnitude of $E_\nu=|{\bf p}_\nu|=\textrm{M}\Lambda$, as
with the decay into two particles.
On the other hand,
if the electron has a specific non-zero kinetic energy, there is a whole family of arrangements for the
3-momenta of the rest two particles. These arrangements are described by the intersection of the ellipsoid
with major axis equal to the rest of energy shared by the other two particles
($m_n-m_e-T_e$) and $\textrm{B},\textrm{K}$ as foci, with the momenta screen.
The maximum allowed energy for the electron is such that there is but a unique arrangement for the other two particles' momenta.
In order to have a single intersection point (unique solution)
of an ellipsoid with a plane on which one of the ellipsoid's foci is lying, the ellipsoid should be a degenerate one
with major axis equal to the focal distance.
This is the case where no energy is left for the antineutrino; therefore it is exactly
the situation with only two particles, an electron and a proton, which was analyzed above. The whole range of
energies for the electron within this interval of kinetic energies are actually observed in beta-decay experiments.

\begin{figure}[h]
\begin{center}
\scalebox{0.60}{\includegraphics{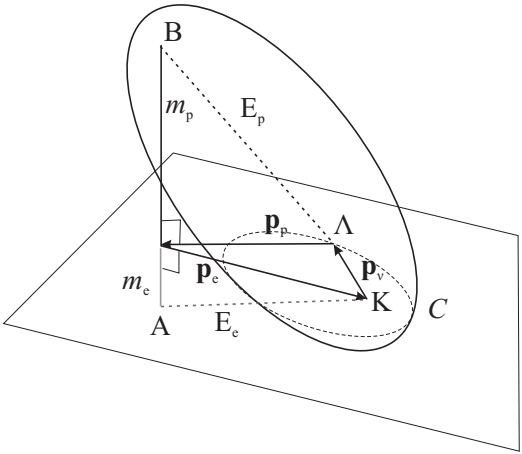}}
\caption{\label{fig:18} The arrangements of ${\bf p}_\nu$ and ${\bf
p}_p$ are described by the intersection of an ellipsoid with the
plane of momenta screen, as with problem 4.}
\end{center}
\end{figure}

\begin{figure}[h]
\begin{center}
%\scalebox{0.50}{\includegraphics{betad1.eps}}
\centerline{\includegraphics[width=18pc] {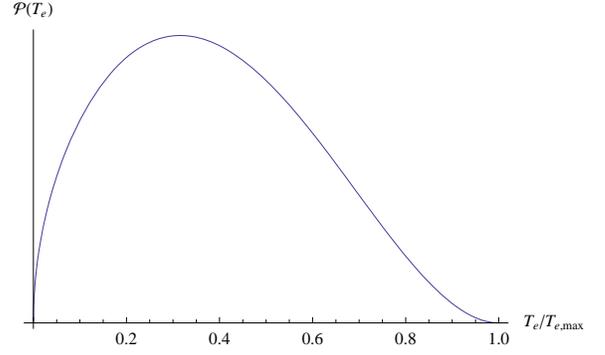}}
\caption{\label{fig:19} The probability distribution of $T_e$ in
beta decay due to kinematics only.}
\end{center}
\end{figure}

Besides predicting the range of the spectrum itself, we can also
predict the distribution function for the energy of the electron, at
least with respect to kinematics. For a given kinetic energy of the
electron, which corresponds to a magnitude of $|{\bf p}_e|$, there
is a whole family of vector arrangements for the 3-momenta of the
rest two particles. In order to visualize these arrangements, we
should draw the ellipsoid of constant energy
$E_p+E_\nu=m_n-E_e=m_n-m_e-T_e$ (this  is the major axis of the
ellipsoid), with the ending point of ${\bf  p}_e$, \textrm{K}, and
the free edge of the vertical segment $m_p$, \textrm{B}, as its
foci, and then find its intersection with the plane of momenta
screen. The ellipse $\it{C}$ (see Fig.~\ref{fig:18}) that will arise
from such an intersection is the locus of all possible ${\bf p}_\nu$
that are compatible with the particular value of $T_e$. Therefore
the possible arrangements of the three particles' 3-momenta that
correspond to $T_e$ for the electron, are described by the surface
of a sphere of radius $|{\bf  p}_e|$ (all possible directions for
the electron), times the surface of the ellipsoid that comes about
when the ellipse $\it{C}$ is rotated around the direction of ${\bf
p}_e$ (all possible inclinations of the plane of momenta around the
axis of ${\bf  p}_e$). Hence, the distribution function of $T_e$
will be given by the product of the two surfaces. The surface of the
corresponding sphere is \be S^{\textrm{sph}}=4 \pi |{\bf  p}_e|^2
\ee while the surface of an ellipsoidal surface that arises from the
revolution of $\it{C}$ around ${\bf p}_e$ is \be S^{\textrm{ell}}=2
\pi a b \left(
\frac{b}{a}+\frac{\sin^{-1}\sqrt{1-(b/a)^2}}{\sqrt{1-(b/a)^2}}
\right) \ee where $a,b$ are the semi-major and the semi-minor axes
of the ellipse respectively. Finally, the derivative $d|{\bf
p}_e|/dT_e$ is needed to transform the distribution into a
distribution over $T_e$: \be \label{Prob} {\cal P}(T_e)=\frac{d
|{\bf  p}_e | }{dT_e}S^{\textrm{sph}} S^{\textrm{ell}} =4 \pi
(T_e+m_e)|{\bf  p}_e| S^{\textrm{ell}}. \ee Computing the dimensions
$a,b$ of the ellipse is straightforward, though quite tedious. It
can be performed through algebraic computations similar to that used
in problem 4 to obtain the intersection of an ellipsoid with a
plane. In Fig.~\ref{fig:19} we have plotted the spectrum of the
electron as a function of its kinetic energy $T_e$
(Eq.~(\ref{Prob})) for a neutron undergoing beta decay, assuming it
was initially at rest.

It should be noted though, that the usual experimental curves for beta decay correspond to
a  metastable nucleus undergoing beta decay, instead of a single neutron. This affects the value
of $T_{e,\max}$ (the maximum value of $T_e$) and the general shape of the curve. 
Also, our description is accurate with respect only to
the kinematics implied by special relativity, without taking into account the internal dynamics of
the transformation.

%%%%%%%%%%%%%%%%%%%
%%%%%%%%%%%%%%%%%%%

%%%%%%%%%%%%%%%%%%%
%%%%%%%%%%%%%%%%%%%
%\end{document}
%%%%%%%%%%%%%%%%%%%%%%%%%%%%%%%%%%%%%%%%%%%%%%
\section{Conclusion}
\label{sec:9}
%%%%%%%%%%%%%%%%%%%%%%%%%%%%%%%%%%%%%%%%%%%%%%

The graphic tools we have used in this paper are simply right triangles,
suitably drawn in order to describe the kinematics of either simple or more complicated
relativistic reactions.

%%%%%%%%%%%%%%%%%%%
\begin{figure}[h]
\begin{center}
\scalebox{0.70}{\includegraphics{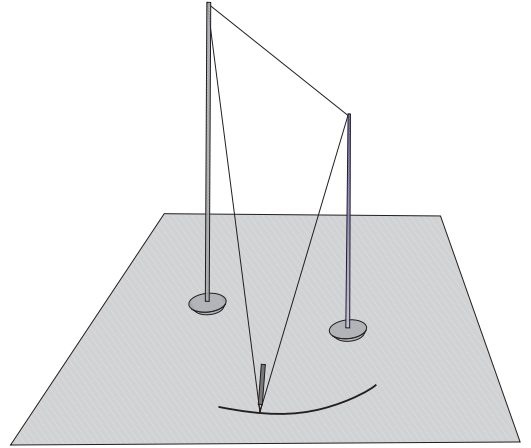}}
\caption{\label{fig:20} A mechanical construction to draw right
triangles corresponding to relativistic particles that take part in
relativistic reactions.}
\end{center}
\end{figure}
%%%%%%%%%%%%%%%%%%%

Using these tools when teaching relativistic reactions, not only is fun,
but also renders complicated analytic computations unnecessary. The student can easily
visualize the kinematics allowed by energy and momentum conservation. The teacher can benefit
from using 3-D diagrams with right triangles, when she or he attempts to construct new problems with reactions.
After a quick drawing she or he could make sure that the problem has the desired solution.
Furthermore, the teacher could make teaching more vivid by using
a simple mechanical construction: a simple metal planar board, rods of adjustable length to use as mass segments with
suitable magnetic bases so that one could stick them perpendicular to the board, an adjustable string and a board marker
(see Fig.~\ref{fig:20})
to build 3-D diagrams that correspond to any relativistic reaction one may think of. We bet that such
a construction will make the analysis of relativistic kinematics even more fun.

%%%%%%%%%%%%%%%%%%%%%%%%%%%%%%%%%%%%%%%%%%%%%%
%{Acknowledgements}
%%%%%%%%%%%%%%%%%%%%%%%%%%%%%%%%%%%%%%%%%%%%%%
\begin{acknowledgments}
This work was supported  by
the research funding program ``Kapodistrias'' with Grant
No 70/4/7672.
\end{acknowledgments}

%%%%%%%%%%%%%%%%%%%%%%%%%%%%%%%%%%%%%%%%%%%%%%

\end{document}